# The Two–Point Correlation Function and the Size of Voids

Dalia S. Goldwirth[1], Luiz Nicolaci da Costa[2,3] and Rien Van de Weygaert[4,5]
[1] *School of Physics and Astronomy, Faculty of Exact Sciences, Tel-Aviv University, 69976 Tel-Aviv, Israel*
[2] *Institut d'Astrophysique 98 bis Boulevard Arago, F75014, Paris, France*
[3] *Observatório Nacional, Rua Gen. José Cristino 77, Rio De Janeiro, Brazil*
[4] *CITA, University of Toronto, 60 St. George St., M5S 1A7 Toronto, Canada*
[5] *Max-Planck-Institut für Astrophysik, Karl-Schwarzschild-Straße 1, Postfach 15 23, 85740 Garching bei München, Germany*

26 February 1995

**ABSTRACT**

Under the assumption of a void-filled Universe we investigate if the characteristic scale of voids can be determined from existing surveys. We use the Voronoi tessellation to create mock surveys and study the properties of the first zero-crossing of the two-point correlation function for various survey geometries. Our main conclusion is that the available data sets should be able to discriminate between 5000 kms$^{-1}$ and 12000 kms$^{-1}$ voids, if one of these scales actually characterizes the distribution of galaxies.

## 1 INTRODUCTION

The data accumulated by complete and dense surveys of the nearby Universe strongly suggest that galaxies tend to lie along thin wall-like structures surrounding large empty regions (Geller & Huchra 1989, da Costa et. al. 1994). Moreover, voids with typical size of about 5000 kms$^{-1}$ appear to be relatively common features of the galaxy distribution as they are seen in different directions of the sky. Deeper surveys of a cross-section of the galaxy distribution out to about 50000 kms$^{-1}$, like the Las Campanas Survey (Schectman et. al. 1992), lend further support to these findings. The qualitative picture that emerges is of a close-packed, volume-filling network of voids. Deep pencil-beam surveys also seem to agree with this cellular-like picture except that they detect structures on a scale of about 128 $h^{-1}$ Mpc (Broadhurst et. al. 1990).

Despite the growing evidence that large voids and Great-Wall like features are common very little has been done in terms of quantifying their frequency and distribution of sizes, primarily due to the intrinsic difficulty of identifying voids as well-defined entities. On the other hand, the existence of large voids or, if one prefers, extended and thin coherent wall-like structures surrounding large empty regions is puzzling and their frequency may pose a serious challenge to theories of structure formation. To our knowledge, with the exception of a few case studies (e.g. Zeng & White 1991), there has not been a systematic detailed study of this question within the context of N-body simulations.

An attempt to quantitatively understand the constraints that the existence of large voids may impose on theories of structure formation was made by Blumenthal et. al. (1992) and Piran et. al. (1993), assuming that the present-day voids are the result of the evolution of proto-voids. According to this scenario the voids seen in the galaxy distribution are a natural consequence of the gravitational evolution of primordial underdense regions which expand in time. Numerical experiments (Regős & Geller 1991, Dubinski et. al. 1993, Van de Weygaert & Van Kampen 1993, Einasto et. al. 1994, Sahni et. al. 1994), together with simple theoretical and statistical arguments (Blumenthal et. al. 1992), suggest that the evolution of underdense regions may lead to a natural explanation for the geometrical properties of the matter and galaxy distribution. The models suggest that the galaxy distribution would be characterized by the size (diameter) of voids reaching shell-crossing today. These voids would also be the largest ones, filling a large fraction of the volume, which would be surrounded by high-density contrast walls (Dubinski et. al. 1992).

The similarity between the predicted distribution in these models and the observed galaxy distribution underscores the need for a more detailed understanding of the nature of the galaxy distribution. In particular, one would like to know how to determine whether the observed galaxy distribution is in fact void-filled, whether there is a characteristic scale for the observed voids and how successful different cosmogonies are in producing the abundance of observed voids.

Unfortunately, until now no suitable statistics exist to address all aspects of these issues. In contrast to overdensities, which collapse to form small bound entities, voids are large and difficult to characterize in a point distribution. High-order statistics like the VPF (e.g. Lachieze-Rey et. al. 1992), and variants (Vogeley 1993 and references therein), have been used in the past. However, these statistics are sensitive to shot-noise, probe relatively small scales and are not convenient for comparing different samples as they depend on the mean density of the sample (see e.g. Little & Weinberg 1993). Similar problems are present in other statistics like nearest-neighbor distribution (Ryden and Turner 1984). The most useful void statistic available at present appears to be the void spectrum analysis developed by Kauffmann & Fairall (1991), with which they managed to identify a large number of voids in the galaxy distribution, and which was shown to be a good indicator of the characteristic void "size" in N-body simulations (Kauffmann & Melott 1992). However, the method makes a priori assumptions about void topology. The genus analysis of the smoothed density field, introduced by Gott, Weinberg and Melott (1987), might shed more light on that issue. However, so far no conclusive results have been obtained, although some interesting conclusions could be drawn on the structure of voids in gravitational instability scenarios when combining the topology analysis with the void spectrum (Sahni, Sathyaprakash & Shandarin 1994).

Here we explore the possibility of using the two-point correlation function $\xi(r)$ for determining the maximum linear extent of voids or diameters in case of spherical voids. The reason for returning to the two-point correlation function is that it is a simple statistic, well-suited for a point process, independent of topology or other a priori assumptions, while for a cellular like distribution the first-zero crossing is a direct measure of the characteristic size of the cells. For instance, for a cubic lattice with a cell size L the correlation function can be expressed analytically and the first zero-crossing occurs at L/2 (e.g. Heavens 1985, also see Van de Weygaert, 1991, ch. 4). Furthermore, preliminary calculations utilizing data from several different new surveys suggest that the scale for zero-crossing is close to 25 $h^{-1}$ Mpc, consistent with voids of 5000 kms$^{-1}$ in diameter. Unfortunately, it is well-known that $\xi(r)$ is very uncertain on large scales due to sampling fluctuations and uncertainties in the adopted normalization. The question we investigate is how large these uncertainties are for a given survey geometry and, considering the errors, if the first zero-crossing can be used to discriminate between different scales.

In this investigation we make extensive use of the Voronoi tessellation which is physically motivated and should be at least a fair representation of the skeleton of the LSS. We utilize a Monte Carlo approach using the Voronoi tessellation as the underlying distribution to determine the errors in $\xi(r)$ and the statistical properties of the first zero-crossing from data subject to selection effects and finite sampling.

In section 2 we describe the toy model we have used in our Monte Carlo experiments. In section 3 we investigate some specific examples of surveys. Our conclusions are summarized in section 4.

## 2  MONTE CARLO SIMULATIONS

### 2.1  Model for the LSS

Most theories of the formation of structure on large scales are based on the gravitational instability scenario (e.g. Peebles 1980). In such scenarios voids are a natural consequence of the evolution of primordial underdense regions. Underdense regions expand rapidly while overdense regions collapse (Icke 1984, but see Babul & Van de Weygaert 1994). Thus, eventually the large underdense regions will become the dominant morphological and, in many cases, dynamical component of the Universe. This crucial role of voids was already stressed more than a decade ago by Zel'dovich, Einasto & Shandarin (1982). Studies of the gravitational growth of proto-voids lead to a scenario of a void-filled universe with voids having a characteristic scale set by those voids shell-crossing today (Blumenthal et al. 1992; Dubinski et al. 1993). Therefore, it seems reasonable to construct a model of the mass distribution in the Universe based on the evolution of the low-density regions. The results of the numerical experiments (Regös & Geller 1991, Dubinski et. al. 1993, Van de Weygaert & Van Kampen 1993, Little & Weinberg 1993, Einasto et. al. 1994, Sahni et. al. 1994) suggest that matter should flow away from the centers off the proto-voids until it encounters similar material flowing out of an adjacent proto-void. If, at a given epoch, the excess expansion in the dominant shell-crossing voids is rather similar, an assumption supported by the study of Dubinski et. al. (1993), matter will collect on planes that perpendicularly bisect the axes connecting the expansion centers. This process leads to the formation of a geometrical structure which resembles a Voronoi tessellation. Based on these considerations we here assume that the skeleton of the matter distribution is approximately represented by a Voronoi tessellation (Van de Weygaert 1991b, 1994). The same physical motivation has led several authors to use this distribution to model the LSS produced in the explosive scenario (e.g. Yoshioka & Ikeuchi 1989).

The use of Voronoi tessellations has recently become quite popular. After the earlier work of Kiang (1966) and Matsuda & Shima (1984), the statistical properties of 2-d foams were extensively examined by Icke & Van de Weygaert (1987). Subsequently, Van de Weygaert & Icke (1989) developed a geometrical 3-D Voronoi algorithm as well as an early version of a code to distribute particles within the walls, edges and vertices of the Voronoi network according to a heuristic prescription of the evolution of the large scale matter distribution. Amongst others, they found that the two-point correlation function of Voronoi vertices has a slope and amplitude that is strikingly similar to the one displayed by rich Abell clusters. Since then there has been an increased interest in the use of Voronoi tessellations as a useful description of the large scale structure (Coles 1990; Van de Weygaert 1991a,b; Ikeuchi & Turner 1991; SubbaRao & Szalay 1992; Williams 1992; Williams et al. 1991). The main virtue of the Voronoi foam is that it provides a conceptually simple model for a cellular or foamlike distribution of galaxies, whose ease and versatility of construction makes it an ideal tool for statistical studies. Its usefulness in this context is underlined by the fact that it is physically motivated and resembles the observed distribution composed of sheets, fil-aments, and clusters surrounding voids. Although the model cannot say much about the pattern of the galaxy distribution on small scales, it is nevertheless a useful prescription for the spatial distribution of the walls themselves.

One might argue that the ideal approach to calculating errors in $\xi$ would be to do a large series of N-body simulation of a model with power spectrum well matched by the real data, and use them to define errors and covariances in the correlation function. However, this approach is computationally very demanding and not suitable for our purposes since it is model dependent and we have no a priori knowledge of the scale of voids. On the other hand, Voronoi tessellations can be constructed relatively fast and are easy to manipulate, so that the behavior of relevant parameters can be studied systematically in a more efficient way. For example, the scale of voids is an input parameter.

### 2.2 Mock Surveys

Formally a Voronoi tessellation is defined as a set of regions $\Pi_i$ of nucleus $i$. For a nuclei with coordinates $X_i$ the region $\Pi_i$ is defined as the set of all points which are nearer to $X_i$ than to any other $X_j$, $j \neq i$. Hence, Voronoi regions are convex polyhedra (3-D) with finite size.

We developed a new version of a numerical Voronoi particle code to construct a galaxy distribution in a 3-D periodic box. A random distribution of expansion centers is the starting point of this procedure. Their number density $n_s$ of nuclei determines the mean size of the voids, which is given by $d_s = 2(3/4\pi)^{1/3} n_s^{-1/3}$.

We restrict ourselves to a Poisson distribution of seeds. This is a plausible approximation if the present-day voids evolved from density minima in a Gaussian random phase density fluctuation field. The voids that have evolved to the shell-crossing phase at the present epoch correspond to density minima on scales where the initial correlations were not very large or even negligible. Although we also could have considered correlated and anticorrelated seed distributions, we chose not to do so. Firstly, for moderate to negligible seed clustering the correlation properties of points in Voronoi walls and filaments, as well as of Voronoi vertices, differs only marginally from that in Poisson Voronoi tessellations (Van de Weygaert 1991b). Furthermore, it is important to emphasize that even in the case of Poisson seeds there will be a distribution of void sizes. Consideration of correlated or anticorrelated seed distributions would introduce an additional scale into the problem, which for correlated seeds would lead to a broadening of the distribution of voids' sizes in the Voronoi foam and as consequence the distribution of zero-crossings. We believe that in order to properly address the main issue of whether $\xi$ can pick up a characteristic scale in a non-regular void distribution when errors of finite sampling and survey geometry are taken into account it is essential to investigate simple models. Consideration of a correlated seed distributions would only unduly complicate the problem without necessarily being a more realistic model.

Subsequently, the galaxies are inserted into the network, their number density being determined by the galaxy luminosity function (see below). The position of each galaxy is determined as follows. A point is inserted at a random position in the simulation box, and we determine the closest expansion center to this point. By definition the inserted point is situated within the Voronoi cell of this 'nucleus'. Subsequently, we determine geometrically the position, closest to the nucleus, on the line passing through the nucleus and the point that is as close to this nucleus as to some other expansion center in the box. Evidently, this position is within the bisecting plane, and more specifically the Voronoi wall, between the two expansion centers. We put a galaxy at this position in the wall. By repeating this procedure for the required number of points, each wall of the Voronoi polyhedra gets filled with a random distribution of galaxies, each wall having a uniform surface number density whose value is linearly proportional to the distance between the defining expansion centers. Notice that this also implies the absence of power on small scales.

Although the code has been developed to allow for more sophisticated cases, by introducing more realistic distributions in which the vertices of the tessellations are identified with clusters, and the edges of the walls are identified with filaments, in most of our simulations we limited ourselves to the simple case in which the galaxies are randomly distributed on the walls. By an additional displacement of each of the galaxies perpendicular to their walls, the amplitude of the displacement determined by sampling from a distribution function (in our case Gaussian or uniform), the walls are given a finite thickness. We have adopted a typical wall thickness of 500 kms$^{-1}$. Tests have indicated that our conclusions are not sensitive to the thickness of the walls.

In most of our Monte Carlo experiments we adopted a single (average) void scale. We have used voids with typical sizes of 25, 50 and 100 $h^{-1}$ Mpc, which cover the range of scales of interest. A more realistic scenario for the universe, as pointed out by several authors (Dubinski et al. 1993, Van de Weygaert 1991b, Van de Weygaert & Van Kampen 1993, Sahni et. al. 1994), would be a hierarchy of void sizes. Although, in principle we could attempt to simulate such structure the lack of dynamics in our construction would lead to an artificial distribution of expansion centers and therefore to unrealistic models. Moreover, we are primarily interested in the typical scale of the void distribution which we expect will also have the most prominent ridges.

In order to simulate real observations galaxies were assigned magnitude according to a Schechter luminosity function given by $\varphi$

$$\varphi = \varphi^* \left(\frac{L}{L^*}\right)^\alpha e^{-L/L^*} . \qquad (1)$$

In our calculations the parameters for the luminosity function were taken, whenever possible, from the determination of the original authors. Only galaxies brighter than $M_{min} = -17$ were considered since as we discuss below we consider only semi-volume limited samples in the calculation of the two-point correlation function. This bright cutoff was also necessary for the simulation of the deep surveys, as the number of galaxies increases dramatically.

Our Monte Carlo simulations consist of choosing a typical void size, generating a tessellation in a box of size appropriate for the depth of the survey considered and assigning magnitudes according to equation (1). Assuming that the observer is located in the middle of the box we extract for each Voronoi tessellation a sample with the same geometry, effective and apparent magnitude limits as the survey considered. We used 1000 realizations for each case we studied. In figure 1 we show a typical cut of such a tessellation with cells of a characteristic size of 50 $h^{-1}$ Mpc using the same selection criteria as those of the SSRS2 survey (see section 3.1).

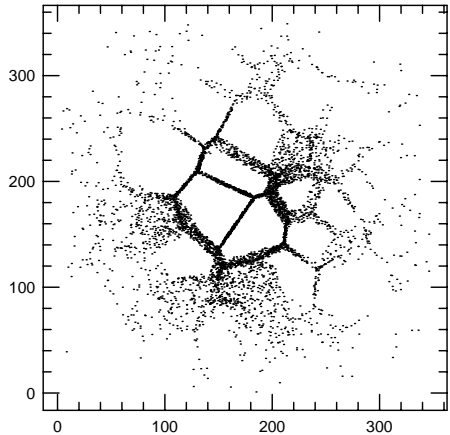

**Figure 1.** A 20 $h^{-1}$ Mpc thick cut of a Voronoi tessellation in a 3-d box of 350 $h^{-1}$ Mpc to the side. The galaxies were selected using the same selections as in the SSRS2 survey (see section 3.1).

### 2.3 The Two-point Correlation Function

We define $\xi(r)$ as the probability in excess of Poisson distribution of finding a galaxy in a volume $\delta V$ a distance r, away from a randomly chosen galaxy,

$$\delta P = n \delta V \left[1 + \xi(r)\right] \quad , \tag{2}$$

where n is the mean number density of galaxies.

In order to calculate $\xi$ we generate a random sample with the same selection criteria as that of the Voronoi tessellation. We then use the estimator

$$\xi(r) = \frac{N_{DD}(r)}{N_{DR}(r)} \frac{n_R}{n_D} - 1 \tag{3}$$

where $N_{DD}$ and $N_{DR}$ refer to the number of data-data and data-random pairs, respectively, at a separation $r \pm \delta r$ ($\delta r \ll r$). This estimator is unaffected by the geometry of the slices since the random sample has the same geometry and suffers from the same boundary effects. The estimator in Equation (3) can be generalized to include weighting functions. The weight can depend both on the distance of the objects from the origin, and the distance of the objects from each other. We chose to weigh the number of pairs, $N$, and the mean number density of the galaxies, $n$, by the selection function $\phi(r)$. We define $\phi(r)$ as

$$\phi(r) = \frac{\int_{L(r)}^{\infty} \varphi(L) dL}{\int_{Lmin}^{\infty} \varphi(L) dL} \tag{4}$$

where $L(r)$ is the minimal observed luminosity at radius $r$ and $Lmin$ is the luminosity that corresponds to $M_{min}$. Specifically we use

$$N_{xy} = \sum \frac{1}{\phi(r_i)\phi(r_j)} \tag{5}$$

where $xy$ stands for $DD$ or $DR$ and the sum is over all the galaxies separated by a distance $r \pm \delta r$. The density $n$ was estimated as

$$n = \frac{1}{V} \sum_i \frac{1}{\phi(r_i)} \tag{6}$$

where V is the volume of the survey (e.g. Davis and Huchra 1982).

We have also performed some calculations using the estimator

$$\xi(r) = \frac{N_{DD} N_{RR}}{N_{DR}^2} - 1 \tag{7}$$

proposed by Hamilton (1993). This estimator is expected to be less sensitive to uncertainties in the mean density.

## 3 RESULTS

Utilizing the mock surveys we determine the errors in the correlation function due to finite sampling, sampling fluctuations and finite volume. We also examine the distribution of the first zero-crossing of the correlation for the different surveys considered below. As illustrations we have considered in our study the following examples: dense, wide-angle surveys like the combined CfA2 and SSRS2 (Geller and Huchra 1989, da Costa et. al. 1994), moderately deep but sparse surveys like the Stromlo-APM Survey (Loveday et. al. 1992) and deep, thin and almost complete surveys like the Las Campanas Survey (Schectman et. al. 1992). We have chosen these surveys because they are either complete or near completion and represent the range of strategies recently adopted by different authors. We have not included in our analysis the deep pencil beam surveys like that of Broadhurst et. al. (1990) since their geometry is not as easily accommodated to our existing code.

In all the calculations of the correlation function presented below we have semi-volume limited the samples analyzed, considering only galaxies with absolute magnitudes brighter than M = -18.5. We have also only considered galaxies within a maximum distance where the selection function drops to about 0.1. Hereafter we refer to that distance as being the "effective" depth of the survey.

### 3.1 Surveys

In our analysis we have not attempted to reproduce exactly the regions and all the details of the selection criteria adopted in the surveys considered but only the more general characteristics like the approximate sky coverage and limiting magnitude.

To represent the geometry of the nearby dense surveys we take the geometry of the combined CfA2 north + south samples, the recently completed SSRS2 of the southern galactic cap and the ongoing survey of the northern galactic cap south of $\delta \leq 0°$. This sample consists of about 18000 galaxies brighter than $m_{B(0)} = 15.5$ covering roughly 1/3 of the sky to an effective depth of about 130 $h^{-1}$ Mpc. The missing chunk of sky in the northern galactic cap should be completed in the near future. Below we refer to a geometry like the SSRS2 (South) as S, with a volume of $8 \times 10^5 h^{-3}$ Mpc$^3$ and a density of $5 \times 10^{-3}$ galaxies $h^3$ Mpc$^{-3}$, and the combined sample SSRS2 (South+North) and CfA2 (South+North) as SNSN, with the same density but 4 times the volume.

Although the parameters describing the Schechter luminosity functions for the CfA2 and the SSRS2 have been found to differ (Marzke et. al. 1994, da Costa et. al. 1994) we have for our purposes adopted a single luminosity function, using the parameters appropriate for the SSRS2 survey with $M^* = -19.6$, $\alpha = -1.2$ and $\varphi^* = 0.97 \times 10^{-2}$ $h^3$ Mpc$^{-3}$.

As a second example we consider the Stromlo-APM Survey (Loveday et. al. 1992, hereafter APM), which randomly selects galaxies at a rate of 1-in-20 drawn from a complete magnitude-limited sample in the range $15 < b_J < 17.15$. This survey covers an area of 4300 square degrees and extends out to about 40000 kms$^{-1}$. The Voronoi tessellations for this case were generated in a 3-D box 800 $h^{-1}$ Mpc to the side, The parameters for the Schechter luminosity function were taken to be $M^* = -19.5$, $\alpha = -0.97$ and $\varphi^* = 1.4 \times 10^{-2}$ $h^3$ Mpc$^{-3}$, as determined by Loveday et. al. (1992). During the process of generating the tessellation galaxies were randomly discarded at the same sampling rate adopted in this survey. In the calculation of the correlation function we further limited the sample to a distance of about 290 $h^{-1}$ Mpc. This survey has a volume of about $10^7 h^{-3}$ Mpc$^3$ and a density of $1.7 \times 10^{-4}$ galaxies $h^3$ Mpc$^{-3}$, about 30 times less than the dense surveys.

As a final example we consider the Las Campanas Survey (Schectman et. al. 1992, hereafter LC) of galaxies in the magnitude range $17.5 < m_{B(0)} < 19.2$ in thin slices. This survey was designed to probe scales where one would expect the distribution of galaxies to reach homogeneity. The survey extends out to 60000 kms$^{-1}$, probing several structures with characteristic sizes as large as 100 $h^{-1}$ Mpc. In the present analysis we only consider the geometry of a thin but completely filled-in strip $120° \times 1.5°$ as the one currently completed. Since the luminosity function for this survey is not available we have adopted the Schechter parameters derived for the SSRS2. In the calculation of the correlation function the maximum distance considered is of about 600 $h^{-1}$ Mpc. In our analysis we neglect other known selection criteria such as the cut in surface brightness and the incompleteness of the survey in dense regions due to the finite number of fibers available for each field. The volume of this survey is roughly the same as that of SNSN with about the same density of galaxies.

### 3.2 Discussion

In figure 2 we present in the different panels the mean correlation function and the errors derived from 1000 Monte

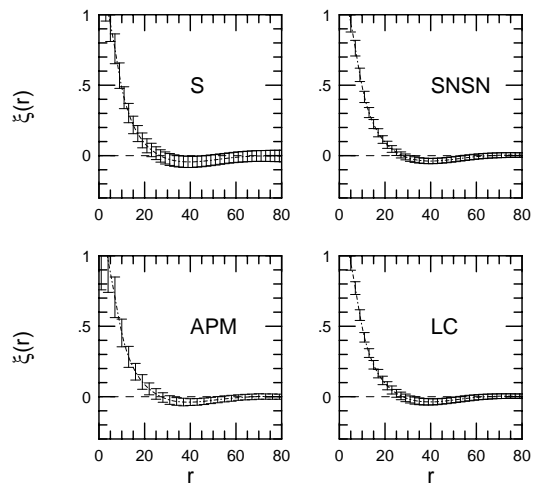

**Figure 2.** For each of the geometries described in the text we show the mean correlation function derived from 1000 semi-volume limited samples. In this example the diameter of the voids is 50 $h^{-1}$ Mpc.

Carlo simulations for each of the surveys described above with voids of size 50 $h^{-1}$ Mpc. As expected the mean correlation obtained from 1000 realizations agrees remarkably well with the all-sky "sample" and with the true correlation function computed from the 3-D Voronoi tessellation free of luminosity and geometrical effects. In particular, the first zero-crossing of the mean correlation function occurs at roughly half of the scale of the typical void imposed in the simulation. Although we only show the case for voids with a typical diameter of $50 h^{-1}$ Mpc, similar results were obtained for the other void sizes.

The Monte Carlo simulations allow us to estimate the amplitude of the error we may expect in the determination of the correlation function from real data due to finite sampling. The errors were computed from the variance of the correlation function at each separation obtained from the 1000 realizations considered and includes contributions from shot-noise, sampling fluctuations as well as other systematic errors due to the finite volume considered. The amplitude of the error, represented by the error bars in figure 2, is shown in figure 3 for the different surveys considered and the different scales of inhomogeneities. In all cases the error is relatively large and comparable to the amplitude of the correlation function in the region where it becomes negative (anti-correlation), especially for large voids (100 $h^{-1}$ Mpc). It has been for this reason that the two-point correlation function has been discarded as a useful statistic to characterize the clustering properties on large scales. Note, however, that this depends on the size of voids and the ge-ometry of the survey. For voids less than 50 $h^{-1}$ Mpc all geometries lead to comparably small errors. On the other hand, for larger voids the error is significantly larger and only the APM survey seems to be suitable to measure the zero-crossing with some degree of confidence (see discussion below).

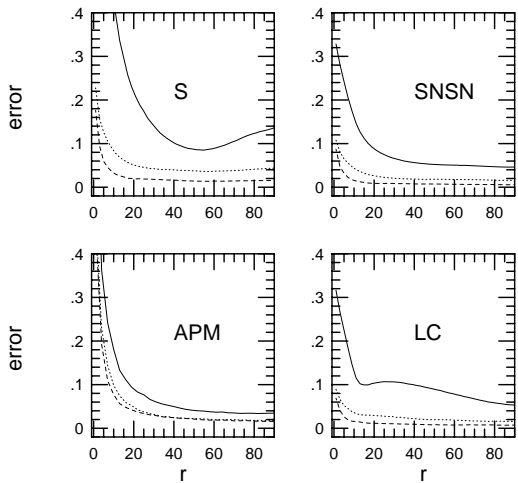

**Figure 3.** The amplitude of the errors in the mean correlation function for each of the geometries described in the text. With 25 $h^{-1}$ Mpc voids (dashed line), 50 $h^{-1}$ Mpc voids (dotted line), 100 $h^{-1}$ Mpc voids (solid line).

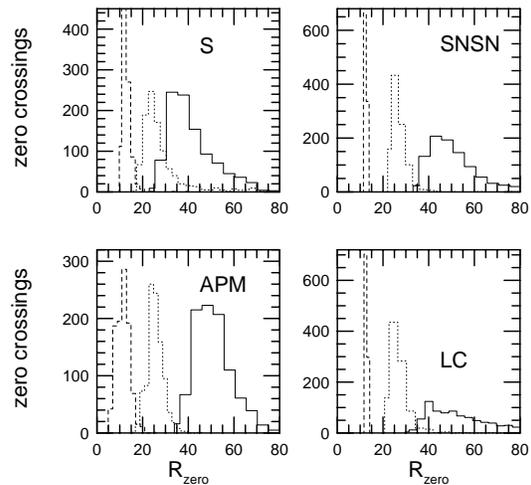

**Figure 4.** The distribution of the first zero crossing for each of the geometries described in the text with 25 $h^{-1}$ Mpc voids (dashed line), 50 $h^{-1}$ Mpc voids (dotted line), 100 $h^{-1}$ Mpc voids (solid line). For each case the distribution is from 1000 simulations.

The ability to detect the scale of the voids from the correlation function as measured from the data should depend on the probability distribution of the first zero-crossing taking into account the various sources of errors. To investigate in more detail the statistical properties of this quantity we have computed the distribution of zero-crossings expected for Voronoi tessellations with typical void sizes of 25, 50 and 100 $h^{-1}$ Mpc. The resulting distributions, obtained from 1000 realizations for the different survey geometries, are shown in figure 4. The effect of the void size relative to the total volume can be clearly seen as the distribution for small voids is relatively well defined and close to a Gaussian. As we increase the void size the distribution broadens and overlaps the distribution obtained for larger voids. In some cases, the maximum of the distribution also shifts to smaller values leading to an underestimate of the true void size. This is a noticeable effect for surveys with small volumes such as the case shown in the upper-left panel for voids 100 $h^{-1}$ Mpc in diameter.

We expect that the effectiveness of a given survey to determine the typical size of voids from the correlation function depends primarily on how well defined the distribution of the zero-crossings are, for the different void scales, and the relative overlap between them. Using this as a qualitative measure we immediately see that both the dense, wide-angle, nearby survey (SNSN) and the APM survey are more suitable to detect the scale of voids than the LC survey. Since the SNSN and LC have approximately the same volume the behavior of the zero-crossing distribution indicates that the critical characteristic of the survey is not only its volume but also its solid angle, which combined determine how many independent voids are fully contained in a given survey. A two-dimensional survey, like the LC survey, although deep, picks up different scales as only a cross-section of the distribution is seen and different parts of the voids are seen (see figure 1). This is a direct consequence of stereological considerations (see e.g. Stoyan, Kendall & Mecke 1987) from which we know that the size of cells in infinitely thin 2-D cross-sections through 3-D tessellations is smaller than the size of the corresponding 3-D cells. In particular it can be shown that the cell volume distribution in a 3-D Voronoi tessellation has a considerably different character than the size distribution of the 2-D cells in a cross-section through that tessellation (Van de Weygaert 1991b, 1994). While the 3-D cell volume distribution is essentially a broad peak centered on the average (typical) cell size, the sectional cell area displays an extended distribution with an average size significantly lower than the average size of its 3-D counterparts (by a factor of $\approx 0.8$), a long tail towards large sizes, and an important and increasing contribution towards small section sizes. Dependent on the survey geometry, and in particular the solid angle they occupy on the sky, the surveys discussed in this paper exhibit behavior that lies somewhere in between the extremes of a full three-dimensional survey and an infinitely thin cross-section. It is clear that the 2-D surveys are less sensitive to the true scale of the voids, and that a survey with a larger solid angle is better suited to constrain the size of voids.

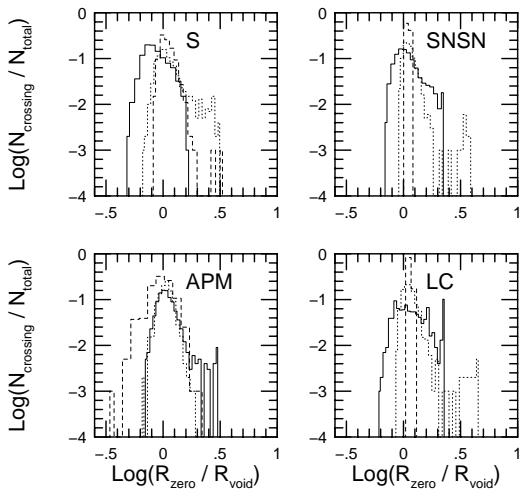

**Figure 5.** The distribution of the first zero crossing as a function of x. For each of the geometries described in the text with 25 $h^{-1}$ Mpc voids (dashed line), 50 $h^{-1}$ Mpc voids (dotted line), 100 $h^{-1}$ Mpc voids (solid line). For each case the distribution is from 1000 simulations.

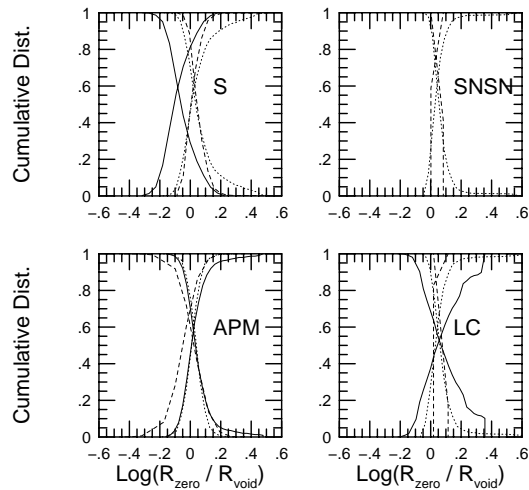

**Figure 6.** The cumulative distribution of the first zero crossing as a function of x. For each of the geometries described in the text with 25 $h^{-1}$ Mpc voids (dashed line), 50 $h^{-1}$ Mpc voids (dotted line), 100 $h^{-1}$ Mpc voids (solid line). For each case the distribution is from 1000 simulations.

To better understand the nature of the distributions shown in figure 4 we present them in a more convenient form in figure 5, where we plot the frequency distribution $f = N_{crossing}/N_{total}$ as a function of $x = log(R_{zero}/R_{void})$ to eliminate the scaling factor and allow a direct comparison between them. Here $R_{zero}$ is the scale of the zero-crossing and $R_{void}$ is the radius of the typical void. From the figure it is clear that surveys probing small volumes (case S) the zero-crossing distribution of large voids is shifted to smaller scale, which leads to a systematic underestimate of their sizes. It is also clear that for the SNSN the distribution is somewhat skewed to large zero-crossings, while APM and LC are more symmetric, the latter being, however, broader then the rest. It is interesting that for the APM the distribution of zero-crossing for the small voids is significantly broader than in the case of more dense surveys. A possible explanation for this is that for a total fixed number of galaxies, a distribution consisting of small voids will have too few points per structure causing the small structures to be poorly delineated yielding a broader distribution. This is apparently the main limitation of a sparse survey.

In order to compute confidence intervals we calculate the cumulative distributions $P_1$ and $P_2$ defined as

$$P_1(x < x_1) = \int_{-\infty}^{x_1} f(x')dx' \qquad (8)$$

and

$$P_2(x > x_2) = \int_{x_2}^{\infty} f(x')dx' \qquad (9)$$

which measure the probability for x to be less than a specified value $x_1$ and for x to be larger than $x_2$, respectively. In figure 6 we display $P_1$ and $P_2$ for all the curves of figure 5.

From the curves shown in figure 6 we can estimate the 95% confidence interval for the zero-crossing scale given the void size and geometry. This was done by determining the values for $x_1$ and $x_2$ for which $P_1$ and $P_2$ were equal to 0.025. From these values we derive the intervals given in table 1. From the table we see that in all cases the samples are sufficiently large to discriminate between voids of 25 and 50 $h^{-1}$ Mpc. However, only the APM and marginally the SNSN can be used to discriminate between 50 and 100 $h^{-1}$ Mpc voids at this confidence level.

Table 1: 95% Confidence Interval for the Zero-crossing Scale.

|  | S | SNSN | APM | LC |
|---|---|---|---|---|
| Void size |  |  |  |  |
| 25 | 10.4-18.3 | 11.9-15.5 | 6.6-18.8 | 12.1-16 |
| 50 | 18.9-66.6 | 22.73-40.5 | 20.5-35.4 | 22-57.7 |
| 100 | 28.8-71 | 37.4-102.4 | 38.5-104.4 | 37-157 |

The same analysis was performed using the Hamilton (1993) estimator defined above for the nearby surveys S and SNSN. In general, we find that the zero-crossing distribution is essentially unchanged except for the large zero-crossing tail which is not as extended, leading to a slightly narrower intervals than those listed in table 1. A major difference was found only for the case of voids 50 $h^{-1}$ Mpc in the S case. This is probably because this is the most sensitive to the uncertainties of the mean density. Still we believe that for comparison with data this estimator should be preferred.

We can also use these probability distributions in a slightly different way and ask, given a measurement of $R_{zero}$, what are the lower and upper limits for the size of typical voids. This can be done by assuming that the cumulative distributions $P_1$ and $P_2$ are always, for a given geometry, similar to those obtained for the 25 $h^{-1}$ Mpc and the 100 $h^{-1}$ Mpc voids, respectively. With this assumption and given a measured value for $R_{zero}$ we estimate the 95% confidence interval for the typical diameter of voids for each geometry considered. These intervals are presented in table 2.

Table 2: 95% Confidence Interval for $D_{void}/R_{zero}$

| S | SNSN | APM | LC |
|---|---|---|---|
| 1.37-3.51 | 1.61-2.67 | 1.33-2.60 | 1.56-2.70 |

For the SSRS2, for instance, preliminary results suggest $R_{zero} \approx 38$ $h^{-1}$ Mpc which, from table 2, implies that the diameter of the typical void is in the range 52 to 133 $h^{-1}$ Mpc at a 95% confidence level. Unfortunately, the range is sufficiently large not to allow to discriminate between the voids seen in the nearby surveys and those claimed by Broadhurst et. al. (1990) from deep pencil-beam surveys. However, if a similar value is obtained for the other surveys we would be able to set some useful constraints on the largest possible diameter for voids. In particular, if the measured zero-crossing is less than about 45 $h^{-1}$ Mpc we would be able to reject at a 95% confidence level the existence of typical voids 128 $h^{-1}$ Mpc in diameter. We hope that these considerations will serve as an incentive for the different groups to publish the values obtained for the zero-crossing. We point out that the zero-crossing may be slightly more difficult to measure in combined samples like the SSRS2 and CfA2 because of possible problems of non-uniformity in the selection criteria. This is still being investigated and we hope to be able to report on that in a future paper. We point out that the zero-crossing may be more difficult to measure in combined samples like the SSRS2 and CfA2 because of possible problems of non-uniformity in the selection criteria and because of effects introduced by redshift distortions. This is still being investigated and we hope to be able to report on that in a future paper.

## 4  CONCLUSIONS

In this paper we have investigated the possibility of determining the typical scale of voids from a simple statistic given by the first zero-crossing of the two-point correlation function. The nice feature of using the properties of the two-point correlation function is that it is simple and adequate to describe a point distribution as it does not impose any scale as other statistics do. It also has a very clear signature for regions devoid of galaxies.

We have derived the probability distribution obeyed by this quantity for different redshift surveys using a Monte Carlo technique which should incorporate random as well as systematic errors to which the data is normally subjected. Our main conclusion is that despite the large errors in the determination of the correlation function on large scales ( > 20 $h^{-1}$ Mpc), the zero-crossing statistic may be a useful tool in determining the scale of typical voids under the assumptions that: 1) the galaxy distribution is void-filled; 2) that there is a characteristic scale for the void distribution; 3) that the Voronoi tessellation is a suitable model for describing the gross features of the large scale structure. These assumptions although reasonable and expected from the visual inspection of the current redshift maps are by no means definite. In particular, we have not proven that the universe is void-filled. However, the existence of large coherent features and the fact that a large fraction of the observed volume is empty argues in favor of this "hypothesis'. We also note that for comparing our results to real data one must take into account the effects of redshift distortions and a valid comparison may require calculating the real space rather than the redshift space correlation function.

Our most important conclusion is that the existing surveys are sufficiently large to allow us to detect the scale of typical voids in a void-filled Universe. In particular, the Stromlo-APM survey should provide us with a good test to check the claims of Broadhurst et. al. (1990) for the existence of voids as large as 128 $h^{-1}$ Mpc.

**Acknowledgments**

We would like to thank W. Press for his helpful suggestions and his generous offer of allowing us to use his computer for the extensive calculations required. We would also like to thank Tsvi Piran for helpful suggestions and comments. LNdC and DG are grateful for the hospitality of the Harvard-Smithsonian Center for Astrophysics. RvdW also would like to thank V. Icke and B. Jones for useful suggestions and discussions while developing the Voronoi particle code. One of us (LNdC) is also thankful to the John S. Guggenheim Foundation for its support and to the hospitality of the Hebrew University of Jerusalem, where part of this work was carried out.

The Two–Point Correlation Function and the Size of Voids    9